\documentstyle[letter,epsf]{ptptex}

\title{ A Scale Invariance due to Compositeness Condition 
	in the Induced Gauge Theory}

\author{
Akira {\sc Akabane}\footnote{ E-mail: akabane@saitama-med.ac.jp} 
and Keiichi {\sc Akama}\footnote{
E-mail: akama@saitama-med.ac.jp}
}

\inst{
Department of Physics, Saitama Medical College,
  Kawakado, Moroyama, Saitama, 350-04, Japan 
}

\recdate{
}

\abst{
The scale invariance of the coupling constant in the induced gauge theory due to its compositeness condition is demonstrated in the renormalization group flow of the finite-cutoff gauge theory at the leading order in $1/N$,
	where $N$ is the number of the matter fermion species.
}

\begin{document}
\maketitle


From theoretical points of view,
	the gauge fields can be dynamically induced as composite
	even without preparing them 
	as fundamental objects.
The idea of the composite gauge bosons 
	are applied to QED \cite{QED}, hadron physics \cite{Had},
	the electroweak theory \cite{EW,TCA,EW2}, 
	QCD \cite{TCA,QCD}, induced gravity \cite{IG}, braneworld \cite{Emb} 
	the theory of hidden local symmetry \cite{Hid} etc.
Some people observe that gauge structures emerge in  
	connection with quantization 
	on non-trivial sub-manifolds \cite{QM2}.
Gauge fields are also induced 
	associated with geometric (Berry) phases \cite{GP}
	which appear 
	in the optical, molecular, solid state, nuclear, cosmological 
	and various physical systems \cite{GP2,GP3,Kik}.

In these models, we expect that 
	the gauge fields become dynamical propagating fields
	through the quantum fluctuations of the matters.
Unfortunately these models are not renormalizable.
A powerful way to analyze such models is to rely on their equivalence
	to the ordinary gauge theory 
	under the compositeness condition (CC), $Z_3=0$ \cite{CC},	
	where $Z_3$ is the wave-function renormalization constant 
	of the gauge field.
In recent years, Hattori and one of the present authors (K.\ A.)
	investigated theoretical aspects of the CC, 
	and observed a complementarity between gauge boson compositeness
	and asymptotic freedom of the theory \cite{CCAH}.
The renormalization group (RG) approaches provide important clues 
	to investigate these model.
In fact many people attempted to study and apply the method \cite{RG}.
Most of them, however, considered the limit of the infinite momentum cutoff, 
	which necessitates some additional assumptions,
	such as existence of fixed point and ladder approximation.
Here, we take the momentum cutoff 
	as a large but finite physical parameter, 
	and make no further assumptions.
In the previous paper, we analyzed the RG flow of the Yukawa model
	and the scale invariance of the coupling constants
	in the Nambu-Jona-Lasinio model \cite{beta}.
In this letter, we show scale invariance of the coupling constant 
	in the induced gauge theory due to its compositeness condition
	at the leading order in $1/N$, 
	where $N$ is the number of the fermion species.
We demonstrate it in the renormalization group flow 
	of the finite-cutoff gauge theory 
	which includes the induced gauge theory as a special case.

We consider the induced (composite) gauge fields $V_\mu$
	described by the Lagrangian 
\begin{eqnarray} 
	{\cal L}_{\rm comp}=
	\overline {\Psi}(i\!\not\!\partial-\not\! V - M)\Psi
\label{Lcomp}
\end{eqnarray} 
where $\Psi=(\Psi _{1}, \Psi _{2}, \cdots , \Psi _{N})$ 
	are $N$ fermions, 
	$M$ is the mass of $\Psi$.
Classically $V_\mu$ is not an independent dynamical variable
	since it has no kinetic term.
Its Euler equation, however, imposes a constraint, 
	giving rise to interactions among $\Psi$, 
	which supply the kinetic term of $V_\mu$
	through their quantum fluctuations.
Thus $V_\mu$ is interpreted as a quantum composite field.
The form of the Lagrangian (\ref{Lcomp}) is the common essential step
	in the various theories of the induced (composite) gauge field
	\cite{QED,Had,EW,TCA,EW2,QCD,IG,Emb,Hid,QM2,GP,GP2,GP3,Kik}.
We want to show that the coupling constant $g$ is independent of the scale.

In order to see it, 
we consider the Abelian gauge theory  
	for the gauge boson $A^0_\mu$ and $N$ matter fermions 
	$\psi^0=(\psi^0_{1},\psi^0_{2},\cdots,\psi^0_{N})$ 
	with the following Lagrangian
\begin{eqnarray} 
	{\cal L}_{\rm G}=
	-{1\over4}(\partial_\mu A^0_\nu-\partial_\nu A^0_\mu)^2
	+\overline {\psi^0}(i\!\not\!\partial-g_0\not\!\! A^0 - m_0 )\psi^0
\label{LY}
\end{eqnarray} 
where $m_0$ is the bare mass of $\psi^0$, and
	$g_0$ is the bare coupling constant. 
We adopt the dimensional regularization where we consider everything 
	in $d(=4-2\varepsilon )$ dimensional spacetime 
	with small but non-vanishing $\varepsilon $.
By them we are not considering "the theory at the $d(=4-2\varepsilon )$",
	but that at $d=4$ with the momentum cutoff described by the scheme.
The parameter $\varepsilon $ roughly corresponds to $1/\log\Lambda $ 
	with the momentum cutoff $\Lambda $.
To absorb the divergences of the quantum loop diagrams due to (\ref{LY}), 
	we renormalize the fields, the mass, and the coupling constants as
\begin{eqnarray} 
	\psi ^0= \sqrt {Z_2 } \psi ,\ \ \   
	A_\mu ^0= \sqrt {Z_3 } A_\mu ,\ \ \   
	Z_2 m_0^2= Z_m m^2, \ \ \ 
	Z_2 \sqrt {Z_3 }g_0= {Z_1} g \mu ^\varepsilon ,\ \ \ \ \  
\label{ZZ}
\end{eqnarray} 
where $\psi $, $A_\mu $, $m$, and $g$ are the renormalized 
	fields, mass, and coupling constants, respectively, 
	$Z_1$, $Z_2$, $Z_3$, and $Z_m$ are
	the renormalization constants,
	and $\mu $ is a mass scale parameter 
	to make $g$ dimensionless. 
Then the Lagrangian ${\cal L}_{\rm G}$ becomes
\begin{eqnarray} 
	{\cal L}_{\rm G}
	=-{1\over4}Z_3(\partial_\mu A_\nu-\partial_\nu A_\mu)^2
	+Z_2 \overline \psi i\!\not\!\partial \psi 
	+Z_1g\mu ^\varepsilon \overline \psi \not\!\!A_\mu \psi
	+Z_m m\overline \psi \psi. 
\label{LYR}
\end{eqnarray} 
As the renormalization condition, 
	we adopt the minimal subtraction scheme, 
	where, as the divergent part
	to be absorbed into in the renormalization constants, 
	we retain all the negative power terms in the Laurent 
	series in $\varepsilon $ of the divergent (sub)diagrams.
Then the parameter $\mu $ is interpreted as the renormalization scale.
Since the coupling constants are dimensionless,
	the renormalization constants depend on $\mu $ 
	only through $g$, but do not explicitly depend on $\mu $.

We can see that the Lagrangian (\ref{LYR})  
	coincides with the Lagrangian (\ref{Lcomp}) if 
\begin{eqnarray} 
	Z_3 =0,\ \ \ Z_1\not=0,\ \ \ Z_2\not=0.	
\label{CC}
\end{eqnarray} 
The condition (\ref{CC}) is the ``compositeness condition" (CC) \cite{CC}
	which imposes relations among the coupling constant $g$, 
	the mass $m$, and the cutoff parameter $\varepsilon $ 
	in the gauge theory
	so that it reduces to the induced (composite) gauge theory.
The perturbative calculation shows that 
	$g\rightarrow0$ 
	as $\varepsilon\rightarrow0$ at each order,
	and the theory becomes trivial free theory.
Therefore we fix the cutoff $\Lambda=\mu e^{1/\varepsilon}$ at some finite value.
We can read off from (\ref{Lcomp}) and (\ref{LYR}) 
	that the fields 	and masses should be connected by the relations
\begin{eqnarray} 
	\Psi=\sqrt{Z_2}\psi,\ \
	V_\mu={Z_1 g\mu^\varepsilon \over Z_2}A_\mu,\ \
	Z_2 M= Z_m m.  
		\label{Cf}
\end{eqnarray} 
In terms of the bare parameters the CC (\ref{CC}) 
	corresponds to the limit
\begin{eqnarray} 
	g_0\rightarrow \infty. \ \ \ \ 
\label{limbare}
\end{eqnarray} 
This behaviors may look singular at first sight,
	but they are of no harm 
	because they are unobservable bare quantities.

Let us consider RG equation in the gauge theory 
	with special cares on the finite cutoff.
In our case, it amounts to fix $\varepsilon =(4-d)/2$ at some non-vanishing value.
The beta functions and the anomalous dimensions are defined as 
\begin{eqnarray} &&
	\beta _g^{(\varepsilon )}(g)=\mu {\partial g\over \partial \mu },
\label{betadef}\\
&&
	\gamma _{A_\mu} ^{(\varepsilon )}(g)
	={1\over 2}\mu {\partial \ln Z_\phi \over \partial \mu }\ , \ \ \ \ 
	\gamma _\psi ^{(\varepsilon )}(g)
	={1\over 2}\mu {\partial \ln Z_\psi \over \partial \mu }\ ,
\end{eqnarray} 
where the differentiation $\partial /\partial \mu $ performed with
	$g_0$ and $\varepsilon $ fixed.
Operating $\mu (\partial /\partial \mu )$ to the equations in (\ref{ZZ}) we obtain
\begin{eqnarray} 
	\left[ \beta _g^{(\varepsilon )}{\partial \over \partial g}
	+\varepsilon \right] gJ=0, \ \ \ \ 
\end{eqnarray} 
where $J=Z_1/(Z_2 \sqrt {Z_3 })$.
Comparing the residues of the poles at $\varepsilon =0$, we obtain
\begin{eqnarray} 
	\beta _g^{(\varepsilon )}=-\varepsilon g+g{\cal D}J_1
\label{beta}
\end{eqnarray} 
where ${\cal D}=g(\partial /\partial g)$, and 
	$J_1$ is the residue of the simple pole 	of $J$.
On the other hand the anomalous dimensions are given by
\begin{eqnarray} 
	\gamma _{A_\mu} ^{(\varepsilon )}=-{1\over 2}{\cal D}Z_{3, 1}, \ \ \ \ 
	\gamma _\psi ^{(\varepsilon )}=-{1\over 2}{\cal D}Z_{2, 1}, \label{gamma}
\end{eqnarray} 
where $Z_{3, 1}$ and $Z_{2, 1}$ are the residues of the simple poles 
	of $Z_{3}$ and $Z_{2}$, respectively.
We can read off from (\ref{beta}) and (\ref{gamma})
	that $\beta_g ^{(\varepsilon )}$ depends on the cutoff only through
	the first terms $-\varepsilon g$ of the expression,
	while $\gamma ^{(\varepsilon )}$'s are independent of $\varepsilon $.
We should be careful not to neglect the cutoff dependence 
	of $\beta_g ^{(\varepsilon )}$.

The CC $Z_3 =0$ (eq.(\ref{CC})) with (\ref{ZZ1}) 
	connects the terms with different order in the coupling constants.
Accordingly, the expansion in the coupling constants fails 
	in the case of the induced gauge theory.
Therefore we instead adopt the $1/N$ expansion by assigning 
\begin{eqnarray} 
	g^2\sim1/N,\ \ \ \
\label{gsim}
\end{eqnarray} 
which does not mix the different orders in CC (\ref{CC}).
Explicit calculations at the leading order in $1/N$ show
\begin{eqnarray} 
	Z_3 =1-{Ng^2\over 12\pi ^2\varepsilon }\ ,\ \ \ Z_1=Z_2 =1.\ \ \
\label{ZZ1}
\end{eqnarray} 
Applying (\ref{gsim}) and (\ref{ZZ1}) to (\ref{beta}) and (\ref{gamma}), 
	we get, at the leading order in $1/N$,
\begin{eqnarray} 
	\beta _g^{(\varepsilon )}=-\varepsilon g+{Ng^3\over 12\pi ^2},
\ \ \ 
	\gamma _{A_\mu} ^{(\varepsilon )}={Ng^2\over 12\pi ^2}\ ,\ \ \ 
	\gamma _\psi ^{(\varepsilon )}=0.\ \ \  
\label{beta1}
\end{eqnarray} 
The RG equation 
	with the functions in (\ref{beta1}) 
	determine the flow of the various quantities 
	with the increasing scale $\mu$.

The RG equations for the coupling constant $g$ 
	are given by (\ref{betadef}) with (\ref{beta1}),  
	and are solved as
\begin{eqnarray} 
		g^2={ 			1		
\over 
\displaystyle 	{N\over 12\pi ^2\varepsilon }+{\mu ^{2\varepsilon }\over g_0^2}
}\ ,
\label{sol1}
\end{eqnarray} 
where the integration constant has been determined 
	in accordance with (\ref{ZZ}).
We can confirm the results by deriving (\ref{sol1}) 
	directly from (\ref{ZZ}) with (\ref{ZZ1}).
In fact the RG flow of coupling constant 
	is entirely determined by (\ref{ZZ}).
In the infinite cutoff limit $\varepsilon \rightarrow 0$, 
	(\ref{sol1}) becomes
\begin{eqnarray} 
		g^2={		1            
\over 
\displaystyle 		a-{N\ln\mu ^2\over 12\pi ^2} 
}\ , 
\label{sol0}
\end{eqnarray} 
	with $a=N/16\pi ^2\varepsilon +1/g_0^2$  kept fixed.

Let us consider the properties of the induced gauge theory 
	in the RG flow of the cutoff gauge theory.
In the limit of induced gauge theory (\ref{limbare}), 
	the solution (\ref{sol1}) reduces to
\begin{eqnarray} 
	g^2={12\pi ^2\varepsilon \over N}\equiv g^2_{\rm comp},\ \ \ \ 
\label{CCsol1}
\end{eqnarray} 
This can also be derived by directly solving the compositeness condition 
	$Z_3 =0$ in (\ref{CC}) with (\ref{ZZ1}).
The coupling constant (\ref{CCsol1}) for the induced gauge theory
	is independent of the scale parameter $\mu$.
Namely, the induced gauge theory is at the fixed point (\ref{CCsol1}) 
	in the RG flow of the cutoff gauge theory.

\begin{figure}
\epsfysize=4cm
\centerline{\epsfbox{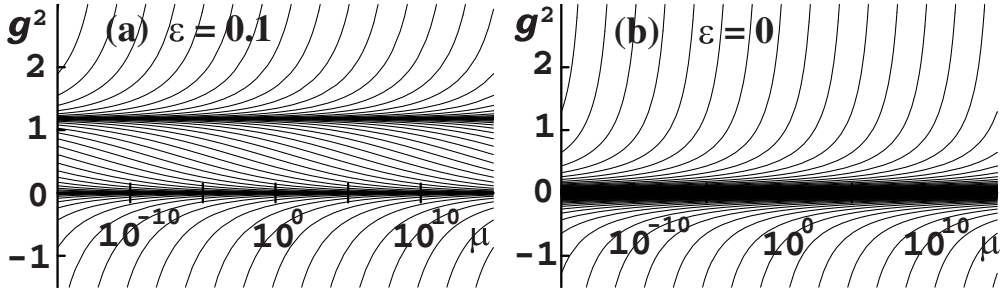}}
\caption{ 
The RG flow of $g^2$ vs $\mu$ at 
(a) finite cutoff and (b) infinite cutoff.
}
\label{f1}
\end{figure}

Fig.\ 1 shows the $\mu$-dependence of $g^2$
	for various values of $g_0^2$,
	(a) at finite cutoff $\varepsilon=0.1$,
	and (b) at the infinite cutoff limit $\varepsilon=0$.
The number of the fermion species $N$ is typically taken as 10.
If $g_0^2>0$, then $0<g^2<g^2_{\rm comp}$, and
	$g^2\rightarrow g^2_{\rm comp}-0$, as $\mu\rightarrow +0$,
	while $g^2\rightarrow +0$ as $\mu\rightarrow \infty$.
If $g_0^2<0$ and 
$0<\mu<(-Ng_0^2/16\pi^2\varepsilon)^{1/2\varepsilon}\equiv\mu_{\rm c}$, 
	then $g^2>g^2_{\rm comp}$ and $g^2\rightarrow g^2_{\rm comp}+0$, 
	as $\mu\rightarrow +0$,
	while $g^2\rightarrow \infty$ as $\mu\rightarrow \mu_{\rm c}-0$.
If $g_0^2<0$ and $\mu>\mu_{\rm c}$, 
	then $g^2<0$, and $g^2\rightarrow -\infty$ 
	as $\mu\rightarrow \mu_{\rm c}+0$,
	while $g^2\rightarrow -0$ as $\mu\rightarrow \infty$.
Thus the $ g^2=g^2_{\rm comp}$ is an infrared fixed point, 
	and $ g^2=0$ is an ultraviolet fixed point.
The region $ g^2<g^2_{\rm comp}$ is asymptotically free,
	though the region $ g^2<0$ is unphysical 
	because the Lagrangian is not hermitian.
On the other hand, in the region $g^2> g^2_{\rm comp}$,
	$g^2$ blows up at $\mu=\mu_{\rm c}$,
	though it should not be taken serious,
	because the expansion itself is no longer justified 
	in the region where $g^2$ is large.
In the infinite cutoff limit $\varepsilon\rightarrow0$,
	the fixed point $ g^2=g^2_{\rm comp}$ 
	moves to and fuses with the other fixed point $ g^2=0$. 
Accordingly the physical asymptotic-freedom region 
	$0<g^2< g^2_{\rm comp}$ disappears, leaving only the region
	where the coupling constant blows up.

Thus the induced gauge theory is at a fixed point in the RG flow
	of the cutoff gauge theory.
The coupling constant in the induced gauge theory is scale-invariant,
	and does not run with the scale parameter.
We can trace back the reason of scale invariance
	to the fact that beta functions vanish 
	due to the compositeness condition.
In fact if we substitute the solution (\ref{CCsol1}) 
	of the compositeness condition,
	the beta function in (\ref{beta1}) vanishes. 
It is further traced back to the fact that the scale invariance of
	the relation  (\ref{ZZ}) in the limit (\ref{limbare}).
Thus the scale invariance is expected to hold not only at 
	the leading order in $1/N$, but also at higher order. 
We are planning to investigate it at the next-to-leading order.

This work is supported by Grant-in-Aid for Scientific Research, 
	Japanese Ministry of Culture and Science.


\begin{thebibliography}{99}

\bibitem{QED}
W.~Heisenberg,			Rev.\ Mod.\ Phys.\ {\bf 29} (1957) 269;
J.~D.~Bjorken,			Ann.\ Phys.\ {\bf 24} (1963) 174;
I.~Bialynicki-Birula, 		Phys.\ Rev.\ {\bf 130} (1963) 465;
D.~Luri\'e and A.~J.~Macfarlane,  Phys.\ Rev.\ {\bf 136} (1964) B816. 
\bibitem{Had}
T.~Eguchi and H.~Sugawara, {Phys.\ Rev.} {\bf D10} (1974) 4257;
H.~Kleinert, Phys.\ Lett.\ {\bf 59B} (1975) 163;
T.~Kugo,  {Prog.\ Theor.\ Phys.} {\bf 55} (1976) 2032;
T.~Kikkawa,  {Prog.\ Theor.\ Phys.} {\bf 56} (1976) 947;
A.~Chakrabarti and B.~Hu, {Phys.\ Rev.} {\bf D13} (1976) 2347.

\bibitem{EW}
P.~Budini,  Lett.\ Nuovo Cim.\ {\bf 9} (1974) 493;
P.~Budini and P.~Furlan,  Nuovo Cim.\ {\bf 30A} (1975) 63;
T.~Saito and K.~Shigemoto, {Prog.\ Theor.\ Phys.} {\bf 57} (1977) 242.

\bibitem{TCA}
H.~Terazawa, Y.~Chikashige and K.~Akama, {Phys.\ Rev.} {\bf D15} (1977) 480;
K.~Akama and T.~Hattori,  Phys.\ Rev.\ {\bf D39} (1989) 1992;
			  		{\bf D40} (1989) 3688;
K.~Akama, T.~Hattori and M.~Yasu\`e, Phys.\ Rev.\ D {\bf 42} (1990), 789;
{\bf 43} (1991) 1702.

\bibitem{EW2}
M.~Suzuki,  Phys.\ Rev.\ {\bf D37} (1988) 210;
S.~Ishida and M.~Sekiguchi, {Prog.\ Theor.\ Phys.} {\bf 86} (1991) 491;
K.~Akama and T.~Hattori, Int.\ J.\ Mod.\ Phys.\ {\bf A9} (1994) 3503;
B.~S.~Balakrishna and K.~T.~Mahanthappa, 
	Phys.\ Rev.\ {\bf D49} (1994) 2653; {\bf D52} (1995) 2379;  
A.~Galli, Phys.\ Rev.\ {\bf D51} (1995) 3876; 
	  Nucl.\ Phys.\ {\bf B435} (1995) 339. 

\bibitem{QCD}
V.~A.~Kazakov and A.~A.~Migdal, Nucl.\ Phys.\ {\bf B397} (1993) 214. 

\bibitem{IG}
A.~D.~Sakharov,  Dokl.\ Akad.\ Nauk SSSR {\bf 177} (1967) 70
	[{Sov.\ Phys.\ Dokl.} {\bf 12} (1968) 1040];
K.~Akama, Y.~Chikashige and T.~Matsuki,
{Prog.\ Theor.\ Phys.} {\bf 59} (1978) 653;
K.~Akama, Y.~Chikashige, T.~Matsuki and H.~Terazawa,
	 {Prog.\ Theor.\ Phys.} {\bf 60} (1978) 868;
K.~Akama,  {Prog.\ Theor.\ Phys.} {\bf 60} (1978) 1900; 
Prog.\ Theor.\ Phys.\  {\bf 61} (1979), 687;
Phys.\ Rev.\ {\bf D24} (1981), 3073; 
S.~L.~Adler, Phys.\ Rev.\ Lett.\ {\bf 44} (1980) 1567;
A.~Zee, Phys.\ Rev.\ {\bf D23} (1981) 858;
D.~Amati and G.~Veneziano, Phys.\ Lett.\ {\bf 105B} (1981) 358;
			Nucl.\ Phys.\ {\bf B240} (1982), 451;
K.~Akama and I.~Oda,	Phys.\ Lett.\ B {\bf 259} (1991), 431;
Nucl.\ Phys.\ B {\bf 397} (1993) 727.

\bibitem{Emb}
K.~Akama, Lect. Notes in Phys.\ {\bf 176} (1983) 267; {Prog.\ Theor.\ Phys.} 
{\bf 78}, 184 (1987); {\bf 79}, 1299 (1988); {\bf 80} (1988) 935; 
hep-th/0307240 (2003);
G.\ Dvali, G.\ Gabadadze, and M.\ Porrati, Phys.\ Lett.\ {\bf B485} (2000) 208; 
K.~Akama and T.~Hattori,Mod.\ Phys.\ Lett.\ A {\bf 15} (2000) 2017.

\bibitem{Hid}
E.~Cremmer and B.~Julia, Phys.\ Lett.\ {\bf 80B} (1978) 48; 
	Nucl.\ Phys.\ {\bf B159} (1979) 141;
M.~Bando, T.~Kugo, S.~Uehara, K.~Yamawaki and T.~Yanagida,
	Phys.\ Rev.\ Lett.\ {\bf 54} (1985) 1215; 

\bibitem{QM2}
N.~P.~Landsman and N.~Linden, Nucl.\ Phys.\ {\bf B365} (1991) 121; 
D.~McMullan and I.~Tsutsui, Phys.\ Lett.\ {\bf B320} (1994) 287; 
		Ann.\ Phys.\ {\bf 237} (1995) 269; 
S.~Tanimura and I.~Tsutsui, Mod.\ Phys.\ Lett.\ {\bf A10} (1995) 2607. 

\bibitem{GP}
S.~Pancharatnam, Proc.\ Indian Acad.\ Sci.\ {\bf 44A} 247;
G.~Herzberg and H.~C.~Longuet-Higgins, 
			Disc.\ Farad.\ Soc.\ {\bf 35} (1963) 77;
C.~A.~Mead and D.~G.~Truhlar, J.\ Chem.\ Phys.\ {\bf 70(05)} (1979) 2284;
M.~V.~Berry, Proc.\ R.\ Soc.\ London Ser.\ {\bf A392} (1984) 45.

\bibitem{GP2}
G.~Delacr\'etaz {\it et al}, Phys.\ Rev.\ Lett.\ {\bf 56} (1986) 2598;
A.~Tomita and R.~Chiao, Phys.\ Rev.\ Lett.\ {\bf 57} (1986) 937;
D.~Suter {\it et al}, Mol.\ Phys.\ {\bf 61} (1987) 1327.
\bibitem{GP3}
A.~Yu.~Smirnov, Phys.\ Lett.\ {\bf B260} (1991) 161;
S.~Forte, Mod.\ Phys.\ Lett.\ {\bf A6} (1991) 3153;
H.~K.~Lee, M.~A.~Nowak, M.~Rho and I.~Zahed,
			Ann.\ Phys.\ {\bf B227} (1993) 175;
Y.~Aharanov {\it et al}, Phys.\ Rev.\ Lett\ {\bf 73} (1994) 918;
A.~Corichi and M.~Pierri, Phys.\ Rev.\ {\bf D51} (1995) 5870.


\bibitem{Kik}
K.~Kikkawa, Phys.\ Lett.\ {\bf B297} (1992) 89; 
T.~Hatsuda and H.~Kuratsuji, UTHEP-286 (1994);
K.~Kikkawa and H.~Tamura, Int.\ J.\ Mod.\ Phys.\ {\bf A10} (1995) 1597. 

\bibitem{CC}
B.~Jouvet, Nuovo Cim. {\bf 5} (1956) 1133; 
M.~T.~Vaughn, R.~Aaron and R.~D.~Amado, {Phys.\ Rev.} {\bf 124} (1961) 1258;
A.~Salam, Nuovo Cim. {\bf 25} (1962) 224; 
S.~Weinberg, {Phys.\ Rev.} {\bf 130} (1963) 776;
B.~W.~Lee, K.~T.~Mahanthappa, I.~S.~Gerstein and M.~L.~Whippman,
			Ann.\ Phys.\ {\bf B28} (1964) 466;
J.~Lemmon and K.~T.~Mahanthappa, Phys.\ Rev.\ {\bf D13} (1976) 2907; 
T.~Eguchi,  {Phys.\ Rev.} {\bf D14} (1976) 2755; {\bf D17} (1978) 611;
H.~Kleinert, in {\it Understanding the fundamental
constituents of matter, proceedings, 1976 Erice Summer School}, 
ed. A. Zichichci (Plenum Publishing Corporation, 1978), 289;
K.~Shizuya,  {Phys.\ Rev.} {\bf D21} (1980) 2327;

\bibitem{CCAH}
K.~Akama,  		{ Phys.\ Rev.\ Lett.} {\bf 76}, 184 (1996);
Nucl.\ Phys.\ A {\bf 629} (1998), 37C;
K.~Akama and T.~Hattori, { Phys.\ Lett.} {\bf B392}, 383 (1997); 
{\bf B445}, 106 (1998); hep-th/0310236 (2003).


\bibitem{RG}
T.~Eguchi, Ref.\ \cite{CC};
K.\ Akama and T.\ Hattori, Ref.\ \cite{TCA};	
J.~Zinn-Justin,  {Nucl.\ Phys.} {\bf B367} (1991) 105; 
D.~Lurei\'e and G.B.~Tupper,  {Phys.\ Rev.} {\bf D47} (1993) 3580;
J.~A.~Gracey, {Phys.\ Lett.} {\bf B308} (1993) 65; {\bf B342} (1995) 297.

\bibitem{beta}
K.~Akama, hep-th/0310183 (2003). 


\end{thebibliography}
\end{document}